# Boltzmann-Informed Probabilities

Yair Neuman[1] and Yochai Cohen[2]

Head, The Functor Lab, Department of Cognitive and Brain Science, Ben-Gurion University of the Negev, Beer-Sheva 84105, Israel,

yneuman@bgu.ac.il

[2]. Gilasio coding, Tel-Aviv, Israel, yohai@gilasio.com

* Corresponding author(s): Yair Neuman (yneuman@bgu.ac.il)

**Abstract**

Traditional interpretations of probability, whether frequentist or subjective, make no reference to the concept of energy. In this paper, we propose that assigning hypothetical energy levels to the outcomes of a random variable can yield improved probability estimates. We apply this Boltzmann-informed approach to the context of sports betting and analyze five seasons of the English Premier League data. It was found that when used to compute the Kelly criterion, Boltzmann-informed probabilities consistently outperform probabilities derived from the original betting odds. These findings demonstrate the value of integrating energy-informed probabilities into studying complex social systems.

**Keywords**: Boltzmann, interdisciplinarity, energy-conceptual, Kelly criterion, soccer

# Boltzmann-Informed Probabilities

## 1. Introduction

Traditional probability models—frequentist or subjective—do not usually incorporate the concept of energy into the measurement of probabilities. However, in its most general conceptual sense, energy appears whenever some work is done. Therefore, energy may be used as a constraint to be considered whenever we seek to estimate the probability distribution of outcomes requiring different energy levels. The idea can be illustrated with respect to a soccer match. In the 2023-2024 season of the Premier League, the Wolves (Home team) played against Manchester City (Away team). The average betting odds were Home = 9.2, Draw = 5.65, and Away = 1.33. Given these betting odds, showing a clear preference for Manchester City, we can ask how much energy is required for each outcome. Let $O_H$, $O_D$, and $O_A$ be the odds for a home win, draw, or away win. The energy levels can then be *heuristically* defined as:

$E_H = O_H/O_A$

$E_A = O_A/O_H$

$E_D = O_D$

In the case of the Wolves against Man City:

$E_H = 9.2/1.33 = 6.917$

$E_A = 1.33/9.2 = 0.144$

$E_D = 5.65$

These energy levels show that the efforts required by the underdog team – the Wolves – to beat its opponent are much higher (6.917 vs 0.144, respectively). In sum, we can quantify the efforts required for each outcome by translating the betting odds into energy levels/states, which have a relative meaning only.

Through Boltzmann's seminal work, we know that the probability of a state exponentially and inversely decays with respect to its energy level. Defining the energy levels associated with the outcomes of a soccer match, we can compute the Boltzmann distribution and use it to compute the *Dirichlet conjugate prior* to estimate the *posterior* distribution of the possible results of the match. This approach, which follows [1] and [2], is fully explained, detailed, and illustrated in the methodology section. It must be emphasized that we do not naively use the Boltzmann distribution as it appears in its context in physics. We use the Boltzmann distribution as an abstract mathematical modeling tool. This is why we used the term Boltzmann-informed probabilities.

## 2. Boltzmann for betting odds

In statistical mechanics, the Boltzmann distribution models the probability of a system occupying a particular energy state $E$i through the following equation, where k is Boltzmann's constant and $T$ is the temperature:

$$P(Ei) = (e^{-Ei/kT})/Z$$

This equation suggests that lower-energy states are more probable, introducing a natural asymmetry that reflects real-world constraints on the system's behavior. Therefore, by considering the outcomes of a random variable, as representing the probabilities of different energy states, we include real-world constraints in our understanding of a system. For example, in soccer, betting odds are a valuable tool for forecasting the outcome of a soccer match (e.g., [3]). The odds express a large aggregation of information from various sources, from the teams' past performances to injuries of leading players. In this sense, the odds express the outcome of a betting market, and a recurrent question is how efficient this market is. It was argued that "Betting market efficiency implies that market prices (i.e., bookmaker odds) reflect all relevant historical information and represent the best forecasts of the match outcome's probabilities” [4, p. 713. However, the market is not fully efficient, given sources of inefficiency such as behavioral biases (e.g., [5]) or delayed information incorporation (i.e., Markets might not adjust instantly to new information). One known source of bias is the “favourite–longshot bias” [6, p. 803], where low-probability outcomes ("longshots") are systematically overbet and overpriced relative to their actual chances of occurring.

These shortcomings of the betting market can be illustrated by analyzing the 23-24 season of the Premier League. The three possible outcomes of a match are A (Away team wins), D (Draw), and H (Home team wins). The baseline for these outcomes was: 32%, 22%, and 46%, respectively. If one decides to bet on the outcome with the highest probability, then for A, he will be correct in 56% of the cases, and for H, in 61% of the cases. While these hit rates improve prediction over the baseline, they are far from perfect. By reinterpreting odds through the physical lens, we can adjust them to the "effort" or "energy" required by teams to achieve a specific outcome. This approach is introduced in the following sections.

## 3. Methods

We introduce a Boltzmann-informed probabilistic framework to improve the estimation of soccer match outcomes. The method integrates theoretical priors derived from energy-based modeling of betting odds with empirical data from historical match outcomes. It must be emphasized that our focus is not on predicting soccer games. Soccer is used only as an illustrative case. To repeat, we focus on illustrating the potential of Boltzmann-informed probabilities for various modeling tasks. Sport betting is used only as an exemplary case.

### 3.1 Data and preprocessing

We analyzed five consecutive seasons (2019–2023) of the English Premier League. For each match, we extracted:

1. The home and away teams
2. The average betting odds for home win ($O_H$), draw ($O_D$), and away ($O_A$)
3. The match result (home win, draw, or away win)

The following steps describe the procedure used to generate the Boltzmann informed probabilities. The pseudo-code for the methods used in this paper appears in Appendix 1.

**3.2 Step 1. Defining energy levels**

We interpret the odds as proxies for the effort required to achieve a match outcome. The energy levels are then heuristically defined as follows:

$E_H = O_H/O_A$

$E_A = O_A/O_H$

$E_D = O_D$

This formulation reflects the previously discussed intuition that less likely outcomes require greater energy to realize. These energy levels are then transformed into Boltzmann's distribution.

**3.3 Step 2. The Boltzmann probability distribution**

Using the Boltzmann equation:

$$P(Ei) = (e\text{^}{-Ei/kT})/Z$$

with $k = T = 1$ we compute:

$p_H = e\text{^}{-E_H}/Z$

$p_D = e\text{^}{-E_D}/Z$

$p_A = e\text{^}{-E_A}/Z$

where Z is the normalizing function:

$$Z = e^{-E_H} + e^{-E_D} + e^{-E_A}$$

The distribution describes the probability of observing an energy state, equivalent to the match outcome. For example, with respect to the match between the Wolves and Manchester City, the probabilities of winning the game as derived from the betting odds (i.e., the *original* probabilities) are 0.108 for the Wolves and 0.751 for Man City. However, computing the Boltzmann distribution, the probabilities are:

$p_H = 0.001$

$p_D = 0.004$

$p_A = 0.9948$

This means that the probability of the Wolves winning the game has been reduced from 0.10 to 0.001, and the probability of Man City winning the match has increased to 0.9948.

The Boltzmann-informed probabilities are used to estimate the posterior distribution of the match results. Given a distribution of possible match outcomes, as expressed by the betting odds, we assume that our prior beliefs of observing it can be represented by parameters $\alpha = (\alpha_H, \alpha_D, \alpha_A)$. These parameters, derived from the Boltzmann distribution, can be used to form the Dirichlet Distribution, which functions as a conjugate prior to the "real" multinomial distribution of the match results. We hypothesized that by using the Boltzmann-informed probabilities, we can better estimate the distribution of the match results.

The logic and benefits of using the Boltzmann equation are fully discussed. However, here is one possible explanation drawing on the idea of *Max Entropy* [7]. Assume a soccer match with three possible outcomes and their corresponding energy levels:

1. Home win: E(H) = 0.6

2. Draw: E(D) = 0.2 (lowest energy)
3. Away win: E(A) = 1.66 (highest energy)

With these energy levels, the corresponding Boltzmann probabilities would be:

1. $p_H$ = e^(-0.6)/Z = 0.549/Z
2. $p_D$ = e^(-0.2)/Z = 0.819/Z
3. $p_A$= e^(-1.66)/Z = 0.190/Z

Z = 0.549 + 0.819 + 0.190 = 1.558

Therefore:

1. $p_H$ = 0.549/1.558 = 0.352 or 35.2%
2. $p_D$= 0.819/1.558 = 0.526 or 52.6%
3. $p_A$= 0.190/1.558 = 0.122 or 12.2%

These probabilities maximize entropy *for the specific constraint* that the average energy equals: E_avg = 0.352·0.6 + 0.526·0.2 + 0.122·1.66 = 0.211 + 0.105 + 0.203 = 0.519. The max entropy principle suggests the least biased inference possible, based on limited information. More specifically, it suggests that among all the probability distributions that satisfy the known constraints, we should choose the one with the highest entropy. For the betting odds, the maximum entropy principle suggests that among all possible probability distributions with an average energy of 0.519, the Boltzmann distribution (35.2%, 52.6%, 12.2%) has the highest entropy. If we tried to create another distribution with the same average energy (0.519), it would necessarily have lower entropy. Therefore, translating the original betting odds into Boltzmann-informed probabilities has the benefit of identifying the least biased distribution of outcomes under the constraint of energy. These Boltzmann-informed probabilities cannot be used in themselves. They are used to define our prior degrees of belief only. We use them as explained in the next section.

### 3.4 Step 3. Incorporating historical performance

To incorporate prior team performance, we apply the following definitions and procedures. For match $i$ =1 to N. Let $W_H$ and $W_A$ be the counts of previous wins for the home and away teams. Let $D_H$ and $D_A$ be the number of previous draws for the respective teams. Define the draw count as the average:

$$D = (D_H + D_A)/2$$

This gives the historical count vector:

$$C = [W_H, D, W_A]$$

which is used to define a Dirichlet prior with parameters:

$$\alpha_H = p_H \cdot S, \alpha_D = p_D \cdot S, \alpha_A = p_A \cdot S$$

where $S$ = round ($W_H + D + W_A$)

For example, for Wolves against Manchester City, the counts are:

$$C = [W_H = 1, D = 0.5, W_A = 6]$$

With $S = 8$. The alpha parameters are then:

$$\alpha_H = 0.008, \alpha_D = 0.03, \alpha_A = 7.95$$

The Boltzmann distribution is therefore applied to the historical records of the teams up to the specific match, to compute the alpha parameters expressing our prior beliefs in the outcomes.

### 3.5 Step 4. Posterior probability calculation

The final posterior probabilities for the match outcomes are calculated using:

$$Post_H = (W_H + \alpha_H)/T, Post_D = (W_D + \alpha_D)/T, Post_A = (W_A + \alpha_A)/T$$

with

$$T = W_H + D + W_A + \alpha_H + \alpha_D + \alpha_A$$

For our example

$$Post_H = 0.065, Post_D = 0.034, Post_A = 0.90$$

These posterior probabilities combine the market odds, transformed via energy and Boltzmann, with empirical match outcome frequencies to yield new updated predictions.

## 4. Evaluation via Kelly-informed betting simulation

To measure the value of the Boltzmann-informed posterior probabilities, we used a simulation-based testing procedure grounded in the *Kelly criterion*, a well-established method for optimizing bet sizing under uncertainty. This procedure evaluates whether gains derived from the Boltzmann-Dirichlet model are higher than those achieved via the original betting odds. The evaluation process consists of three key stages, and it is applied once to the original probabilities derived from the betting odds and once to the posterior probabilities derived through the Boltzmann-informed calculation. We present the simulation with respect to the posterior probabilities computed through the Boltzmann-informed probabilities.

### 4.1 Step 1: Match outcome prediction

For each match $i$ = 1 to N, we first identify the outcome with the highest posterior probability among the three outcomes: home win ($Post_H$), draw ($Post_D$), or away win ($Post_A$). To predict the match's outcome, we chose the outcome with the highest probability. The selected prediction is then compared to the actual match result (FTR). We define a binary success metric ("hit") for each outcome type:

1. $HIT_H = 1$ if $Post_H$ is maximal and FTR = H; otherwise, $HIT_H = 0$.
2. $HIT_D = 1$ if $Post_D$ is maximal and FTR = D; otherwise, $HIT_D = 0$.
3. $HIT_A = 1$ if $Post_A$ is maximal and FTR = A; otherwise, $HIT_A = 0$.

This step provides a baseline measure of accuracy by tracking whether the model's most likely outcome matched the actual result.

### 4.2 Step 2. The Kelly criterion

Next, we determine the optimal stake for each possible outcome using a fractional Kelly. The Kelly criterion estimates the optimal fraction of capital to bet on each outcome given its expected value:

$$KELLY_O = f \cdot (Post_O - (1 - Post_O / Odds_O))$$

where $O \in \{H, D, A\}$, $Post_O$ is the posterior probability of $O$, $Odds_O$ the corresponding betting odds, and $f = 0.2$. Only positive Kelly values are considered. If $KELLY_O \leq 0$, then no bet is placed.

### 4.3 Step 3. Calculating gains

We computed the gain or loss associated with placing a Kelly informed bet for each match. Gains were calculated as follows:

$$GAIN_O = 100 \cdot KELLY_O \cdot Odds_O$$

If the bet was placed and the prediction was incorrect ($HIT_O = 0$), then:

$$GAIN_O = -100 \cdot KELLY_O$$

The unit 100 assumes a bank role of \$100 for each match. The outcome is the cumulative gain/loss across matches. For comparison, we computed the cumulative gain using the same betting procedure with the probabilities derived from the original betting odds. We hypothesized that if the Boltzmann informed probabilities have an advantage, the gain achieved by using them in the Kelly criterion should be higher than that gained through the original probabilities extracted from the betting odds.

## 4. Results

With the original probabilities, there was no case where the simulation bet on a Draw. Therefore, we define the gains for the Boltzmann-informed probabilities (i.e., GainBoltz) and the gains for the original probabilities (i.e., GainOriginal) by summing the gains placed on Home win and Away win only. Table 1 presents the gains and the edge of using the Boltzmann-informed approach versus the original probabilities.

| Season | GainBoltz | GainOriginal | Edge |
|---|---|---|---|
| **2019-2020** | 1546.07 | 1366.22 | +179.85 |
| **2020-2021** | 1731.42 | 1152.18 | +579.24 |
| **2021-2022** | 1981.67 | 1605.93 | +375.75 |
| **2022-2023** | 1828.51 | 1465 | +363.52 |
| **2023-2024** | 2127.07 | 1783.86 | +343.21 |
| **Sum** | 9214.74 | 7373.17 | +1841.57 |

Table 1. Gains for the two approaches

The results show a consistent benefit of using the Boltzmann-informed probabilities in the Kelly-based betting simulation with an accommodating edge of $1841.57 over the five seasons.

One may argue that the edge of GainBoltz over GainOriginal results from the Bayesian procedures rather than from the Boltzmann-informed probabilities. To test this argument, we run the simulation using the Bayesian procedure on the original probabilities rather than with the Boltzmann-informed probabilities. The gains are presented under the title GainOriginal+Bayes. The results are presented in Table 2.

| Season | GainBoltz | GainOriginal | GainOriginal + Bayes |
|---|---|---|---|
| **2019-2020** | 1546.07 | 1366.22 | 819.21 |
| **2020-2021** | 1731.42 | 1152.18 | 1070.88 |
| **2021-2022** | 1981.67 | 1605.93 | 1134.49 |
| **2022-2023** | 1828.51 | 1465 | 972.29 |
| **2023-2024** | 2127.07 | 1783.86 | 1225.86 |
| **Sum** | 9214.74 | 7373.17 | 5222.73 |

Table 2. Gains comparison across the three procedures

As one can see, the argument should be rejected. The performance of the Bayesian procedure without the Boltzmann-informed probabilities is lower than the rest.

## 5. Discussion

This paper presents a novel application of Boltzmann-informed probabilities to sports betting, showing how integrating energy-based modeling and empirical data improves performance based on the original betting odds. The key empirical finding—that Boltzmann-informed posterior probabilities consistently yielded higher returns when applied within a Kelly criterion betting framework—invites further reflection on the underlying mechanisms driving this improvement. Previously, we discussed the idea that the Boltzmann distribution maximizes the entropy under the constraint of average energy. Here we discuss other aspects of the approach by focusing mainly on its ability to correct biases in betting markets.

The Boltzmann approach treats betting odds not simply as expressions of market-implied likelihoods, but as representing the relative "effort" or "energy" required to realize specific outcomes. By translating odds into energy levels, the framework imposes an asymmetry that mirrors real-world constraints: improbable events require disproportionate energy to be realized. This transforms the raw odds into a probability space that penalizes overoptimistic assessments of underdogs and corrects for market overconfidence in favorites. In other words, the Boltzmann transformation effectively accentuates differences in odds, converting them into energy levels that yield exponentially scaled probabilities. The Boltzmann equation *scales* the original probabilities by considering relative energy constraints (i.e., the efforts required by each team to outperform its opponent) and by performing exponential scaling, resulting in the *amplification of differences*. This exponential scaling emphasizes the certainty in specific match outcomes, correcting for biases resulting from psychological sources.

This sharper discrimination between outcomes is important. While rich in information, betting markets may exhibit probability flattening, especially in cases where uncertainty is high. The Boltzmann transformation amplifies the separation between outcome probabilities, allowing for more decisive predictions. This sharper discrimination is particularly valuable within the Kelly framework, where the size and direction of bets depend critically on the difference between estimated probabilities and market odds.

Behavioral biases—such as overbetting on popular teams or misjudging draw probabilities—can introduce inefficiencies into betting odds. By recasting these odds in an energy-based framework, the Boltzmann method partially filters such noise, producing probability estimates that are more robust and less prone to herd-driven distortions [5]. The result that no bets were placed on draws under the original model, contrasted with the presence of draw bets in the Boltzmann-informed approach, reflects this recalibration of neglected possibilities.

Moreover, the Kelly criterion is highly sensitive to expected and actual values differences. Minor improvements in probability estimation can compound into significant gains across many trials. Because the Boltzmann-informed posterior probabilities better aligned with actual match outcomes, as demonstrated by cumulative gain, allowed for more accurate bet sizing. This synergy between refined probability estimates and Kelly optimization was a central driver of the observed financial edge.

In sum, the approach presented in the paper treats energy, from a conceptual perspective, as a real-world constraint that may be used to calibrate probabilities through the exponential function proposed by Boltzmann. The core idea is that the higher the energy states, the less probable it is and vice versa. In the case of betting odds, the energy levels are derived from market odds, revealing regularities and correcting biases in implied probabilities. The same approach has been applied to armed conflicts [2], showing how to predict whether a country will experience a year with high fatalities. While the current study focused on soccer outcomes as a case example, the broader implications suggest that energy-based probabilistic modeling may prove valuable across various domains where probability, energetic constraints, and uncertainty intersect.

**Appendix 1.**

**PSEUDOCODE 1. Generating the Boltzmann-informed probabilities**

```
For each match i = 1 to N:


    # Step 1: Convert betting odds to probabilities
    PHome[i] = 1 / OddsHome[i]
    PDrew[i] = 1 / OddsDrew[i]
    PAway[i] = 1 / OddsAway[i]


    # Step 2: Define energy levels
    EHome[i] = OddsHome[i] / OddsAway[i]
    EAway[i] = OddsAway[i] / OddsHome[i]
    EDrew[i] = OddsDrew[i]


    # Step 3: Compute Boltzmann probabilities (with k = T = 1)
    expEHome = exp(-EHome[i])
    expEDrew = exp(-EDrew[i])
    expEAway = exp(-EAway[i])
    Z = expEHome + expEDrew + expEAway  # Partition function


    pEHome[i] = expEHome / Z
    pEDrew[i] = expEDrew / Z
    pEAway[i] = expEAway / Z


    # Step 4: Count historical match outcomes
```

```
HomeTeam = matches[i].HomeTeam
AwayTeam = matches[i].AwayTeam

HomeWins = COUNT(matches[1:i-1] WHERE HomeTeam won)
AwayWins = COUNT(matches[1:i-1] WHERE AwayTeam won)

# Average number of draws for both teams
HomeDraws = COUNT(matches[1:i-1] WHERE HomeTeam drew)
AwayDraws = COUNT(matches[1:i-1] WHERE AwayTeam drew)
AverageDraws = (HomeDraws + AwayDraws) / 2

# Create a count vector
C = [HomeWins, AverageDraws, AwayWins]
SumC = SUM(C)
SumC = ROUND(SumC)  # Round to nearest integer

# Step 5: Define alpha parameters
alphaHome = pEHome[i] * SumC
alphaDrew = pEDrew[i] * SumC
alphaAway = pEAway[i] * SumC

# Step 6: Add alphas to historical counts for the Dirichlet distribution
DirichletHome = HomeWins + alphaHome
DirichletDrew = AverageDraws + alphaDrew
DirichletAway = AwayWins + alphaAway
```

# Step 7: Normalize to get posterior probabilities

DirichletSum = DirichletHome + DirichletDrew + DirichletAway

PostHome[i] = DirichletHome / DirichletSum

PostDrew[i] = DirichletDrew / DirichletSum

PostAway[i] = DirichletAway / DirichletSum

**PSEUDOCODE 2. Computing gain through the Kelly criterion**

For Match *i* =1 to N

IF NOT(PostH & PostD & PostA)=0 THEN:

**STEP 1**

Identify the MAX value for (PostH, PostD, PostA)

IF MAX = PostH and FTR = ‘H’ THEN HIT_H = 1

IF MAX = PostH and FTR ≠ ‘H’ THEN HIT_H = 0

IF MAX = PostD and FTR = ‘D’ THEN HIT_D = 1

IF MAX = PostD and FTR ≠ ‘D’ THEN HIT_D = 0

IF MAX = PostA and FTR = ‘A’ THEN HIT_A = 1

IF MAX = PostA and FTR ≠ ‘A’ THEN HIT_A = 0

**Outcome**: Three new columns: HIT_H, HIT_D, HIT_A

**STEP 2.**

KELLY_H = 0.2 * (PostH – $(\frac{1-PostH}{OddsH})$)

KELLY_D = 0.2 * (PostD – $(\frac{1-PostD}{OddsD})$)

KELLY_A = 0.2 * (PostA – $(\frac{1-PostA}{OddsA})$)

**STEP 3**

IF HIT_H = 1 & KELLY_H>0 THEN GAIN_H = (KELLY_H * 100) * OddsH

IF HIT_H = 0 & KELLY_H>0 THEN GAIN_H = (-1) * (KELLY_H*100)

IF HIT_D = 1 & KELLY_D>0 THEN GAIN_D = (KELLY_D * 100) * OddsD

IF HIT_D = 0 KELLY_D>0 THEN GAIN_D = (-1) * (KELLY_D*100)

IF HIT_A = 1 & KELLY_A>0 THEN GAIN_A = (KELLY_A * 100) * OddsA

IF HIT_A = 0 & KELLY_A>0 THEN GAIN_A = (-1) * (KELLY_A*100)

**Outcome**: Three new columns: GAIN_H, GAIN_D, GAIN_A

**PSEUDOCODE 3. Computing gain through the Kelly criterion and the original probabilities**

For Match *i*=1 to N

IF NOT(PHome & PDrew & PAway) = 0 THEN:

**STEP 1**

Identify the MAX value for (PHome, PDrew, PAway)

IF MAX = PHome and FTR = 'H' THEN HIT_H_p = 1

IF MAX = PHome and FTR ≠ 'H' THEN HIT_H_p = 0

IF MAX = PDrew and FTR = 'D' THEN HIT_D_p = 1

IF MAX = PDrew and FTR ≠ 'D' THEN HIT_D_p = 0

IF MAX = PAway and FTR = 'A' THEN HIT_A_p = 1

IF MAX = Paway and FTR ≠ 'A' THEN HIT_A_p = 0

**Outcome:** Three new columns: HIT_H_p, HIT_D_p, HIT_A_p

**STEP 2.**

KELLY_H_p = 0.2 * (PHome – $(\frac{1-\text{PHome}}{OddsH})$)

KELLY_D_p = 0.2 * (PDrew – $(\frac{1-\text{PDrew}}{OddsD})$)

KELLY_A_p = 0.2 * (PAway – $(\frac{1-\text{PAway}}{OddsA})$)

**STEP 3**

IF HIT_H_p = 1 & KELLY_H_p >0 THEN GAIN_H_p = (KELLY_H_p * 100) * OddsH

IF HIT_H_p= 0 & KELLY_H_p >0 THEN GAIN_H_p = (-1) * (KELLY_H_p*100)

IF HIT_D_p = 1 & KELLY_D_p >0 THEN GAIN_D_p = (KELLY_D_p * 100) * OddsD

IF HIT_D_p = 0 & KELLY_D_p >0  THEN GAIN_D_p = (-1) * (KELLY_D_p*100)

IF HIT_A_p = 1 & KELLY_A_p >0 THEN GAIN_A_p = (KELLY_A_p * 100) * OddsA

IF HIT_A_p = 0 & KELLY_A_p >0 THEN GAIN_A_p = (-1) * (KELLY_A_p*100)

**Outcome**: Three new columns: GAIN_H_p, GAIN_D_p, GAIN_A_p